\documentclass[12pt]{article}
\begin{document}
\begin{center}
{\bf Entropy In The Present And  Early Universe: New Small
Parameters And Dark Energy Problem}\\ \vspace{5mm}
A.E.Shalyt-Margolin \footnote{E-mail: a.shalyt@mail.ru; alexm@hep.by}\\
\vspace{5mm} \textit{National Center of Particles and High Energy
Physics, Bogdanovich Str. 153, Minsk 220040, Belarus}
\end{center}
\vspace{0.5cm}
\begin{abstract}
It is demonstrated that entropy and its density play a significant
role in solving the problem of the vacuum energy density
(cosmological constant) of the Universe and hence the dark energy
problem.  Taking this in mind, two most popular models for dark
energy - Holographic Dark Energy Model  and Agegraphic  Dark
Energy  Model - are analyzed. It is shown that the fundamental
quantities in the first of these models  may be expressed in terms
of a new small dimensionless parameter. It is revealed that this
parameter is naturally occurring in High Energy Gravitational
Thermodynamics and Gravitational Holography (UV-limit). On this
basis the possibility of a new approach to the problem of Quantum
Gravity is discussed. Besides, the results obtained on the
uncertainty relation of the pair "cosmological constant - volume
of space-time", where the cosmological constant is a dynamic
quantity, are reconsidered and generalized up  to the Generalized
Uncertainty  Relation
\end{abstract}
\section{Introduction}
The Dark Energy Problem is one of the key problems in a modern
theoretical physics \cite{Dar1}--\cite{Phant1}. The vacuum energy
is still the major candidate to play a role of this energy.
Provided the Dark Energy is actually the vacuum energy, the
indicated problem is reduced to better insight into the essence of
the vacuum energy. This problem has attracted the attention of
researchers fairly recently with understanding that a cosmological
constant determining the vacuum energy density is still nonzero,
despite its smallness. As is known, the cosmological constant
$\Lambda$ has been first introduced in the works of A.Einstein
\cite{Einst} who has used it as a antigravitational term to obtain
solutions for the equations of the General Relativity (GR) in the
stationary case. However, when A.Friedmann has found the solutions
for GR in case of expanding Universe \cite{Fried} and E.Hubble has
derived an extension of the latter, A.Einstein refused from the
cosmological constant considering its introduction to be erroneous
\cite{Pais}.
\\But the situation was not so simple. In \cite{Gliner} it has been
stated that any contribution into the vacuum energy acts exactly
as the cosmological constant $\Lambda$ and the Vacuum Energy
Density is proportional to $\Lambda$.
 The principal
problem of the cosmological constant  resides in the fact that its
experimental value is smaller by a factor of $10^{123}$ than that
derived using a Quantum Field Theory (QFT)
\cite{Zel1},\cite{Wein1}.
\\And the theories actively developed at the present time
(e.g., superstring theory, loop quantum gravity, etc.)offer a
modified quantum theory including, in particular, the fundamental
length at Planck's scale. The estimates of $\Lambda$ obtained on
the basis of these theories may be greatly differing from the
initial ones derived from standard QFT.
\\ In this paper some of the properties of the
Vacuum Energy Density are studied within the scope of a Quantum
Field Theory with UV cutoff (minimal length). Such a theory arises
in the Early Universe in all the models without exception since
the fundamental length (probably on the order of Planck's but not
necessarily) is acknowledged to be of a crucial importance in this
case It is shown that for this case the experimental and
theoretical values are close and may be expressed in terms of a
new small parameter introduced in physics at Planck's scales. Here
some explanation is needed. The point is that a Quantum Field
Theory with minimal length (QFTML) or, what is the same, UV-cutoff
is always originating as a deformation of QFT. This deformation is
understood as an extension of some theory with the use of one or
several additional parameters in such a way that the initial
theory shows itself in the limiting process \cite{Fadd}.  One of
such extensions generated by an additional small dimensionless
parameter, in terms of which the Dark Energy Problem is formulated
and successfully solved, is described in this paper. In so doing
entropy of the Universe and its dynamics play a significant role.
Additionally, within the scope of a dynamic approach to $\Lambda$,
its behavior associated with the Generalized Uncertainty Principle
is studied for the pair "cosmological constant - volume of
space-time". In what follows, there is no differentiation between
the notions of the cosmological constant $\Lambda$ and Vacuum
Energy Density $\rho_{vac}$. Besides, it is demonstrated that a
new small parameter occurs in High Energy Gravitational
Thermodynamics and Gravitational Holography (UV-limit)as well. On
this basis the possibility for a new approach to the problem of
Quantum Gravity is discussed.

\section{Vacuum Energy Density
\\and Most Popular Modern Dark Energy Models}
As noted in Introduction, the Vacuum Energy is a major candidate
for the Dark Energy. At the same time, due to a factor of
$10^{123}$ distinction between the experimental  value
$\rho^{exp}_{vac}$ \cite{Dar1}  and the value $\rho^{QFT}_{vac}$
calculated using standard QFT  \cite{Zel1}--
$\rho^{QFT}_{vac}\approx{m_p^4}$
\begin{eqnarray}
\frac{\rho^{exp}_{vac}}{\rho^{QFT}}\approx10^{-123},
\end{eqnarray}
interpretation of Dark Energy as a Vacuum Energy presents great
difficulties. But there are several methods enabling one to
obviate the difficulties. We can name two most popular and
acknowledged approaches.
\subsection{Holographic Dark Energy Models}
 The basic relation for this model is the "energy" inequality
\cite{CKN}-- \cite{myung2}
\begin{equation} \label{EB}
 E_{\overline{\Lambda}}\leq E_{BH} \to
l^3 \rho_{\overline{\Lambda}}\leq m_{p}^{2}l.
\end{equation}
Here $\rho_{\rm \overline{\Lambda}}=\overline{\Lambda}^{4}$ --
vacuum energy density with the UV cutoff $\overline{\Lambda}$ and
$l$ is the length scale (IR cutoff) of the system. For the
equality in (\ref{EB}) we have the {\bf holographic energy
density}
\begin{equation} \label{hed}
\rho_{\overline{\Lambda}}\sim \frac{m_{p}^{2}}{l^{2}}\sim
\frac{1}{(l_{\rm p}l)^2}.
\end{equation}
Also, from (\ref{EB}) we can get the  "entropic" inequality
(entropy bound)
\begin{equation} \label{ENBa}
S_{\rm \overline{\Lambda}}\le (m_{p}^{2}A)^{3/4},
\end{equation}
where $A=4\pi l^2$ is the area of this system in the spherically
symmetric case.
\\ The number of works devoted to the Holographic Dark Energy Models,
beginning from the first publication ~\cite{CKN}, is ever growing
~\cite{refHDE} to relieve us from citing the whole list.
\subsection{Agegraphic Dark Energy Models}
Agegraphic Dark Energy Models became the subject of study only two
years ago ~\cite{CAI}. These relations were based on the result of
K\'{a}rolyh\'{a}zy for quantum fluctuations of time
\cite{Karo}--\cite{Sas}
\begin{equation}
 \delta t=\lambda t^{2/3}_{\rm p}t^{1/3}.
\end{equation}
Using the uncertainty relation of "energy-time" in the flat space
\begin{equation}
\label{TEU}
\Delta E \sim t^{-1},
\end{equation}
we can obtain the {\bf agegraphic  energy density} \cite{Maz},
\cite{myung2}
\begin{equation}  \label{aed}
 \rho_{\rm \textit{\textbf{T}}}\sim
\frac{\Delta E}{(\delta t)^3}\sim \frac{m_{\rm
p}^2}{\textit{\textbf{T}}^2},
\end{equation}
where $\textit{\textbf{T}}$  is an age of the Universe.
\\The number of publications associated with models of this type is
constantly increasing too ~\cite{refADE}. This is caused by their
relative simplicity and by a sufficiently good coincidence of the
agegraphic energy density $\rho_{\rm \textit{\textbf{T}}}$ with
$\rho^{exp}_{vac}$.
\section{Dark Energy Problem and Quantum Theory with UV Cutoff}
By Holographic Dark Energy Models (explicitly) and by Agegraphic
Dark Energy Models (implicitly) it is implied that QFT, where they
are valid, is actually QFT with the UV cutoff or fundamental
length.
\\As it has been repeatedly demonstrated earlier, a Quantum Mechanics of
the Early Universe (Plank Scale) is a Quantum Mechanics with the
Fundamental Length (QMFL) \cite{Gar1}. The main approach to
framing of QFT with UV cutoff is that associated with the
Generalized Uncertainty Principle (GUP)
\cite{Ven1},\cite{GUPg},\cite{Ahl1} and with the corresponding
Heisenberg algebra deformation produced by this principle
\cite{Magg}--\cite{Nou}.
\\Besides, QMFL has been framed first using the deformed density
matrix and then in the produced corresponding Heisenberg algebra
deformation \cite{shalyt1}--\cite{shalyt10}, the density matrix
deformation $\rho(\alpha)$ in QMFL being a starting object called
the density pro-matrix and the deformation parameter (additional
parameter) $\alpha=l_{min}^{2}/x^{2}$, where $x$ is the measuring
scale and $l_{min}\sim l_{p}$. As indicated in this
paper, the deformation parameter $\alpha$ is varying within the limits
$0<\alpha\leq1/4$. Moreover $\lim\limits_{\alpha\rightarrow
0}\rho(\alpha)=\rho,$ where $\rho$ is the density matrix in the
well-known Quantum Mechanics (QM), and the following condition
must be fulfilled:
\begin{equation}\label{U26aS}
Sp[\rho(\alpha)]-Sp^{2}[\rho(\alpha)]=\alpha+a_{0}\alpha^{2}+...
\end{equation}
The explicit form of the above-mentioned deformation gives an
exponential ansatz:
\begin{equation}\label{U26S}
\rho^{*}(\alpha)=exp(-\alpha)\sum_{i}\omega_{i}|i><i|,
\end{equation}
where all $\omega_{i}>0$ are independent of $\alpha$ and their sum
is equal to 1.
\\ In the corresponding deformed Quantum Theory (denoted as $QFT^{\alpha}$) for average values we have
\begin{equation}\label{U26cS}
<B>_{\alpha}=exp(-\alpha)<B>,
\end{equation}
where $<B>$ - average in well-known QFT
\cite{shalyt6},\cite{shalyt7} denoted as $QFT^{\alpha}$.
 All the variables associated with the considered $\alpha$ -
deformed quantum field theory  are hereinafter marked with the
upper index $^{\alpha}$.
\\ Note that the deformation parameter
$\alpha$ is absolutely naturally represented as a ratio between
the squared UV and IR limits
\begin{equation}\label{U26dS}
\alpha=(\frac{UV}{IR})^{2},
\end{equation}
where UV is fixed and IR is varying.
\\As follows from the
holographic principle \cite{Hooft1}--\cite{Bou3}, maximum entropy
that can be stored within a bounded region $\Re$ in $3$-D space
must be proportional to the value $A(\Re)^{3/4}$, where $A(\Re)$
is the surface area of $\Re$. Of course, this is associated with
the case when the region $\Re$ is not an inner part of a
particular black hole. Provided a physical system contained in
$\Re$ is not bounded by the condition of stability to the
gravitational collapse, i.e. this system is simply non-constrained
gravitationally, then according to the conventional QFT
$S_{\max}(\Re)\sim V(\Re)$, where $V(\Re)$ is the bulk of $\Re$.
However in the Holographic Principle case, as it has been
demonstrated originally by G. 't Hooft \cite{Hooft1} and later by
other authors  (for example R. V. Buniy and S. D. H. Hsu
 \cite{Hsu}), we have
\begin{equation}\label{Hol9}
S_{\max}(\Re) \leq\frac{A(\Re)^{3/4}}{{\l_p}^2}.
\end{equation}
In terms of the deformation parameter $\alpha$, the principal
values of the Vacuum Energy Problem may be simply and clearly
defined. Let us begin with the Schwarzschild black holes, whose
 semiclassical entropy is given by
\begin{equation}\label{D1}
S = \pi {R_{Sch}^2}/ l_p^2=\pi {R_{Sch}^2}
m_p^2=\pi\alpha_{R_{Sch}}^{-1},
\end{equation}
with the assumption that in the formula for $\alpha$ $R_{Sch}=x$
is the measuring scale and $l_p = 1/m_p$. Here $R_{Sch}$ is the
adequate Schwarzschild radius, and $\alpha_{R_{Sch}}$ is the value
of $\alpha$ associated with this radius. Then, as it has been
pointed out in \cite{Bal}, in case the Fischler- Susskind cosmic
holographic conjecture \cite{Sussk1} is valid, the entropy of the
Universe is limited by its "surface"
 measured in Planck units \cite{Bal}:
\begin{equation}\label{D2}
S \leq \frac{A}{4} m_p^2,
\end{equation}
where the surface area $A = 4\pi R^2$ is defined in terms of the
apparent (Hubble) horizon
\begin{equation}\label{D3}
R = \frac{1}{\sqrt{H^2+k/a^2}},
\end{equation}
with curvature $k$  and scale $a$ factors.
\\ Again, interpreting $R$ from (\ref{D3}) as a measuring scale,
we directly obtain(\ref{D2}) in terms of $\alpha$:
\begin{equation}\label{D4}
S \leq \pi\alpha_{R}^{-1},
\end{equation}
where $\alpha_{R}=l^{2}_{p}/R^{2}$. Therefore, the average entropy
density may be found as
\begin{equation}\label{D5}
\frac{S}{V}\leq \frac{\pi \alpha_{R}^{-1}}{V}.
\end{equation}
Using further the reasoning line of \cite{Bal} based on the
results of the  holographic thermodynamics, we can relate the
entropy and energy of a holographic system
\cite{Jac1},\cite{Cai1}. Similarly, in terms of the $\alpha$
parameter one can easily estimate the upper limit for the energy
density of the Universe (denoted here by $\rho_{hol}$)
\cite{shalyt14}:
\begin{equation}\label{D6}
\rho_{hol} \leq \frac{3}{8 \pi R^2} m_p^2 = \frac{3}{8
\pi}\alpha_{R} m_p^4,
\end{equation}
that is drastically differing from the one obtained with well-known
QFT
\begin{equation}\label{D7}
\rho^{QFT}\sim m_p^4.
\end{equation}
Here by $\rho^{QFT}$ we denote the energy vacuum density
calculated from well-known QFT (without UV cutoff) \cite{Zel1}.
Obviously, as $\alpha_{R}$ for $R$ determined by (\ref{D3}) is
very small, actually approximating zero, $\rho_{hol}$ is by
several orders of magnitude smaller than the value expected in QFT
-- $\rho^{QFT}$.
\\ Since $m_p\sim 1/l_p$, the right-hand side of (\ref{D6})
 is actually nothing else but the right-hand side of (\ref{hed})
in Holographic Dark Energy Models (subsection 2.1). Thus, in
Holographic Dark Energy Models the principal quantity, {\bf
holographic energy density} $\rho_{\overline{\Lambda}}$
(\ref{hed}), may be estimated in terms of the deformation parameter $\alpha$.
\\In fact, the upper limit of the right-hand side of (\ref{D6})
is attainable, as it has been demonstrated in \cite{shalyt14} and
indicated in \cite{Bal}. The "overestimation" value of $r$ for the
energy density $\rho^{QFT}$, compared to $\rho_{hol}$, may be
determined as
\begin{equation}\label{D8}
r =\frac{\rho^{QFT}}{\rho_{hol}}=\frac{8 \pi}{3}{\bf
\alpha_{R}^{-1}}
 =\frac{8 \pi}{3} \frac{R^2}{l_p^2}
 =\frac{8 \pi}{3} \frac{S}{S_p},
\end{equation}
where $S_p$ is the entropy of the Plank mass and length for the
Schwarzschild black hole. It is clear that due to smallness of
$\alpha_{R}$ the value of $\alpha_{R}^{-1}$ is on the contrary too
large. It may be easily calculated (e.g., see \cite{Bal})
\begin{equation}\label{D9}
r = 5.44\times 10^{122}
\end{equation}
in a good agreement with the astrophysical data.
\\ Naturally, on the assumption that the vacuum energy density
$\rho_{vac}$ is involved in $\rho$ as a term
\begin{equation}\label{vac1}
\rho = \rho_M + \rho_{vac},
\end{equation}
where $\rho_M$ - average matter  density, in case of $\rho_{vac}$
we can arrive to the same upper limit (right-hand side of the
formula (\ref{D6})) as for $\rho$.
\section{Some Comments on a Dynamic Character of Cosmological
Constant and GUP}
 Generally speaking, $\Lambda$ is referred to as a constant just
because it is such in the equations, where it occurs: Einstein
equations \cite{Einst}. But in the last few years the dominating
point of view has been that $\Lambda$  is actually a dynamic
quantity, now weakly dependent  on time \cite{Ran1}--\cite{Shap}.
It is assumed therewith that, despite the present-day smallness of
$\Lambda$ or even its equality to zero, nothing points to the fact
that this situation was characteristics for the early Universe as
well. Some recent results \cite{Min1}--\cite{Min4} are rather
important pointing to a potentially dynamic character of
$\Lambda$. Specifically, of great interest is the Uncertainty
Principle derived in these works for the pair of conjugate
variables $(\Lambda,V)$:
\begin{equation}\label{CC1}
\Delta\Lambda\, \Delta V \sim \hbar,
\end{equation}
where $\Lambda$ is the vacuum energy density (cosmological
constant). It is a dynamic value fluctuating around zero; $V$ is
the space-time volume. Here the volume of space-time $V$ results
from the Einstein-Hilbert action \cite{Min2}:
\begin{equation}\label{CC2}
S_{EH}\supset \Lambda \int d^{4}x \sqrt{-g}=\Lambda V.
\end{equation}
In this case "the notion of conjugation is well-defined, but
approximate, as implied by the expansion about the static
Fubini--Study metric" (Section 6.1 of \cite{Min1}). Unfortunately,
in the proof per se (\ref{CC1}), relying on the procedure with a
non-linear and non-local Wheeler--de-Witt-like equation of the
background-independent Matrix theory, some unconvincing arguments
are used, making it insufficiently rigorous (Appendix 3 of
\cite{Min1}). But, without doubt, this proof has a significant
result, though failing to clear up the situation.
\\Let us attempt to explain (\ref{CC1})(certainly at an heuristic level)
using simpler and more natural terms involved  with the other,
more well-known, conjugate pair $(E,t)$ - "energy - time". We use
the designations of \cite{Min1},\cite{Min2}. In this way a
four-dimensional volume will be denoted, as previously, by $V$.
\\ Just from the start, the Generalized Uncertainty
Principle (GUP) is used. Then a change over to the Heisenberg
Uncertainty Principle at low energies will be only natural. As is
known, the Uncertainty Principle of Heisenberg at Planck's scales
(energies) may be extended to the Generalized Uncertainty
Principle. To illustrate, for the conjugate pair
"momentum-coordinate" $(p,x)$ this fact has been noted in many works
\cite{Ven1}--\cite{Magg1}:
\begin{equation}\label{CC3}
\triangle x\geq\frac{\hbar}{\triangle p}+\alpha^{\prime}
l_{p}^2\frac{\triangle p}{\hbar}.
\end{equation}
In \cite{shalyt3},\cite{shalyt9} it is demonstrated that the
corresponding Generalized Uncertainty Relation for the pair
"energy - time" may be easily obtained from
\begin{equation}\label{CC4}
\Delta t\geq\frac{\hbar}{\Delta E}+\alpha^{\prime}
t_{p}^2\,\frac{\Delta E}{ \hbar},
\end{equation}
where $l_{p}$ and $t_{p}$ represent Planck length and time, respectively.
\\Now we assume that in the space-time volume $\int d^{4}x \sqrt{-g}=
V$ the temporal and spatial parts may be separated (factored out)
in the explicit form:
\begin{equation}\label{CC5}
V(t)\approx t \overline{V}(t),
\end{equation}
where $\overline{V}$ - spatial part $V$.  For the expanding
Universe such an assumption is quite natural. Then it is obvious
that
\begin{equation}\label{CC6}
\Delta V(t)=\Delta t \overline{V}(t)+ t
\Delta\overline{V}(t)+\Delta t \Delta \overline{V}(t).
\end{equation}
Now we recall that for the inflation Universe the scaling factor
is $a(t)\sim e^{Ht}$. Consequently, $\Delta\overline{V}(t)\sim
\Delta t^{3}f(H)$, where $f(H)$ is a particular function of
Hubble's constant.  From (\ref{CC4}) it follows that
\begin{equation}\label{CC7}
\Delta t\geq t_{min}\sim t_{p}.
\end{equation}
However, it is suggested that, even though $\Delta t$ is
satisfying (\ref{CC7}), its value is sufficiently small in order
that $\Delta V$ be contributed significantly by the terms
containing $\Delta t$ to the power higher than the first. In this
case the main contribution on the right-hand side of (\ref{CC6})
is made by the first term $\Delta t \overline{V}(t)$ only. Then,
multiplying the left- and right-hand sides of (\ref{CC4}) by
$\overline{V}$ , we have
\begin{equation}\label{CC8}
\Delta V\geq\frac{\hbar \overline{V}}{\Delta E}+\alpha^{\prime}
t_{p}^2\,\frac{\Delta E \overline{V}}{ \hbar}= \frac{\hbar}{\Delta
\Lambda}+\alpha^{\prime} t_{p}^2 \overline{V}^{2}\frac{\Delta
\Lambda}{ \hbar}.
\end{equation}
It is not surprising that a solution of the quadratic inequality
(\ref{CC8}) leads to a minimal volume of the space-time
$V_{min}\sim V_{p}=l_{p}^{3}t_{p}$ since (\ref{CC3}) and
(\ref{CC4})  result in minimal length $l_{min}\sim l_{p}$ and
minimal time $t_{min}\sim t_{p}$, respectively.
\\(\ref{CC8}) is of interest from the viewpoint of two limits:
\\1)IR - limit: $t\rightarrow \infty$
\\2)UV - limit: $t\rightarrow t_{min}$.
\\In the case of IR-limit we have large volumes $\overline{V}$  and  $V$ at
low $\Delta\Lambda$. Because of this, the main contribution on the
right-hand side of (\ref{CC8}) is made by the first term, as great
$\overline{V}$ in the second term is damped by small $t_{p}$ and
$\Delta \Lambda$. Thus, we derive at
\begin{equation}\label{CC9}
\lim\limits_{t\rightarrow \infty}\Delta
V\approx\frac{\hbar}{\Delta \Lambda}
\end{equation}
in accordance with (\ref{CC1}) \cite{Min1}. Here, similar to
\cite{Min1}, $\Lambda$ is a dynamic value fluctuating around
zero.
\\And for the case (2) $\Delta\Lambda$ becomes significant
\begin{equation}\label{CC10}
\lim\limits_{t\rightarrow
t_{min}}\overline{V}=\overline{V}_{min}\sim
\overline{V}_{p}=l_{p}^{3}; \lim\limits_{t\rightarrow
t_{min}}V=V_{min}\sim V_{p}=l_{p}^{3}t_{p}.
\end{equation}
As a result, we have
\begin{equation}\label{CC11}
\lim\limits_{t\rightarrow t_{min}}\Delta V = \frac{\hbar}{\Delta
\Lambda}+\alpha_{\Lambda} V_{p}^{2}\frac{\Delta \Lambda}{\hbar},
\end{equation}
where the parameter $\alpha_{\Lambda}$  absorbs all the
above-mentioned proportionality coefficients.
\\ For(\ref{CC11}) $\Delta \Lambda \sim
\Lambda_{p}\equiv\hbar/V_{p}=E_{p}/\overline{V}_{p}$.
\\ It is easily seen that in this case $\Lambda \sim M_{p}^{4}$, in
agreement with the value obtained using a naive  (i.e. without
super-symmetry and the like) quantum field theory
\cite{Wein1},\cite{Zel1}. Despite the fact that $\Lambda $ at
Planck's scales (referred to as $\Lambda(UV) $) (\ref{CC11}) is
also a dynamic quantity, it is not directly related to
well-known $\Lambda $ (\ref{CC1}),(\ref{CC9}) (called $\Lambda(IR)
$) because the latter, as opposed to the first one, is derived
from Einstein's equations
\begin{equation}\label{CC12}
R_{\mu \nu} - \frac{1}{2} g_{\mu \nu} R = 8\pi G_N \left( -
\Lambda g_{\mu \nu} + T_{\mu \nu} \right).
\end{equation}
However, Einstein's equations (\ref{CC12}) are not valid at the
Planck scales and hence $\Lambda(UV) $  may be considered as some
high-energy generalization of the conventional cosmological
constant, leading to $\Lambda(IR) $ in the low-energy limit.
\\In conclusion, it should be noted that the right-hand side of
(\ref{CC3}),(\ref{CC4}) in fact is a series. Of course, a similar
statement is true for (\ref{CC11}) as well.
 \\Then, we obtain a
system of the Generalized Uncertainty Relations for the Early
Universe (Planck's scales) in the symmetric form as follows:
\begin{equation}\label{CC13}
\left\{
\begin{array}{lll}
\Delta x & \geq & \frac{\displaystyle\hbar}{\displaystyle\Delta
p}+ \alpha^{\prime} \left(\frac{\displaystyle\Delta
p}{\displaystyle p_{pl}}\right)\,
\frac{\displaystyle\hbar}{\displaystyle p_{pl}}+... \\ &  &  \\
\Delta t & \geq & \frac{\displaystyle\hbar}{\displaystyle\Delta
E}+\alpha^{\prime} \left(\frac{\displaystyle\Delta
E}{\displaystyle E_{p}}\right)\,
\frac{\displaystyle\hbar}{\displaystyle E_{p}}+...\\
  &  &  \\
  \Delta V & \geq &
  \frac{\displaystyle \hbar}{\displaystyle\Delta \Lambda}+\alpha_{\Lambda}
  \left(\frac{\displaystyle\Delta \Lambda}{\displaystyle \Lambda_{p}}\right)\,
  \frac{\displaystyle \hbar}{\displaystyle \Lambda_{p}}+...
\end{array} \right.
\end{equation}
The latter of relations (\ref{CC13}) may be important when
finding the general form for $\Lambda(UV) $, low-energy limit
$\Lambda(IR) $, and also may be a step in the process of
constructing future quantum-gravity equations, the low-energy
limit of which is represented by Einstein's equations
(\ref{CC12}).
 \\ It should be noted that a system of inequalities (\ref{CC13})
may be complemented by the Generalized Uncertainty Relation in
Thermodynamics \cite{shalyt3},\cite{shalyt9},\cite{shalyt11}. Let
us consider the thermodynamics uncertainty relations between the
inverse temperature and interior energy of a macroscopic ensemble
\begin{equation}\label{CC14}
\Delta \frac{1}{T}\geq\frac{k}{\Delta U},
\end{equation}
where $k$ is the Boltzmann constant. \\ N.Bohr \cite{Bohr} and
W.Heisenberg \cite{Heis-term} have been the first to point out
that such kind of uncertainty principle should take place in
thermodynamics. The thermodynamic uncertainty  relations
(\ref{CC14})  were proved by many authors and in various ways
\cite{Lind}; their validity does not raise any doubts.
Nevertheless, relation (\ref{CC14}) was proved in view of the
standard model of the infinite-capacity heat bath encompassing the
ensemble. But it is obvious from the above inequalities that at
very high energies the capacity of the heat bath can no longer be
assumed infinite at the Planck scale. Indeed, the total energy of
the pair heat bath - ensemble may be arbitrary large but finite
merely as the Universe is born at a finite energy. Hence the
quantity that can be interpreted as a temperature of the ensemble
must have the upper limit and so does its main quadratic
deviation. In other words the quantity $\Delta (1/T)$ must be
bounded from below. But in this case an additional term should be
introduced into (\ref{CC14})
\cite{shalyt3},\cite{shalyt9},\cite{shalyt11}:
\begin{equation}\label{CC14a}
\Delta \frac{1}{T}\geq\frac{k}{\Delta U} + \eta\,\Delta U,
\end{equation}
where $\eta$ is a coefficient. Dimension and symmetry reasons give
$$\eta \sim \frac{k}{E_p^2}\enskip.$$
 As in the previous cases,
inequality (\ref{CC14a}) leads to the fundamental (inverse)
temperature  \cite{shalyt3},\cite{shalyt9},\cite{shalyt11}.
\begin{equation}\label{CC15}
T_{max}=\frac{\hbar}{\Delta t_{min} k}\sim \frac{\hbar}{t_{p} k},
\quad \beta_{min} = {1\over kT_{max}} = \frac{\Delta
t_{min}}{\hbar}.
\end{equation}
In the recently published work \cite {Farmany})   the black hole
horizon temperature has been measured with the use of the Gedanken
experiment. In the process the Generalized Uncertainty Relations
in Thermodynamics (\ref{CC14a}) have been derived also. Expression
(\ref{CC14a}) has been considered in the monograph \cite{Carroll}
within the scope of the mathematical physics methods.
\\ Besides, note that one of the first studies of the cosmological
constant within the scope of the Heisenberg Uncertainty Principle
has been presented in several works \cite{Padm} -- \cite{Padm2}
demonstrating the inference: {\bf "vacuum fluctuation of the
energy density can lead to the observed cosmological constant"
\cite{Padm}}. In these works, however, no consideration has been
given to GUP, whereas UV-cutoff has been derived artificially.
\section{Gravitational Thermodynamics and Gravitational Holography
in Low and High Energy}
In the last decade a number of very
interesting works have been published. We can primary name the
works of T.Padmanbhan \cite{Padm1}--\cite{Padm12}, where
gravitation, at least for the â spaces with horizon, is directly
associated with thermodynamics and the results obtained
demonstrate a holographic character of gravitation. Of the
greatest significance is a pioneer work written by T.Jacobson
\cite{Jac1}. For black holes the association has been first
revealed by Bekenstein and Hawking \cite{Bek1},\cite{Hawk1}, who
related the black-hole event horizon temperature to the surface
gravitation. T.Padmanbhan, in particular in \cite{Padm11}, has
shown that this relation is not accidental and may be generalized
for the spaces with horizon.  As all the foregoing results have
been obtained in a semiclassical approximation, i.e. for
sufficiently low energies, the problem arises: how these results
are modified when going to higher energies. In the context of this
paper, the problem may be stated as follows: since we have some
infra-red (IR) cutoff $\textit{L}$ and ultraviolet (UV) cutoff
$l$, we naturally have a problem how the above-mentioned results
on Gravitational Thermodynamics are changed for
\begin{equation}\label{GT1}
\textit{L}\rightarrow l.
\end{equation}
According to Section 3 of this paper, they should become dependent
on the deformation parameter $\alpha$. After all, in the already
mentioned Section 3 (\ref{U26dS}) $\alpha$ is indicated as nothing
else but
\begin{equation}\label{GT2}
\alpha=\frac{l^{2}}{\textit{L}^{2}}.
\end{equation}
In fact, in several papers \cite{acs}--\cite{Kim1} it has been
demonstrated that thermodynamics and statistical mechanics of
black holes in the presence of GUP (i.e. at high energies) should
be modified.  To illustrate, in \cite{Park} the Hawking
temperature modification has been computed in the asymptotically
flat space in this case in particular. It is easily seen that in
this case the deformation parameter $\alpha$ arises naturally.
Indeed, modification of the Hawking temperature is of the
following form(formula (10) in \cite{Park}):
\begin{equation}\label{GT3}
T_{GUP}=(\frac{d-3}{4\pi})\frac{\hbar r_{+}}{2\alpha^{\prime
2}l^{2}_{p}}[1-(1-\frac{4\alpha^{\prime 2}l_{p}^{2}}{
r_{+}^{2}})^{1/2}],
\end{equation}
where $d$ is the space-time dimension, and $r_+$ is the
uncertainty in the emitted particle position by the Hawking
effect, expressed as
\begin{equation}\label{GT4}
\Delta x_i \approx r_+
\end{equation}
and being nothing else but a radius of the event horizon;
$\alpha^{\prime}$ -- dimensionless constant from GUP. But as we have
$2\alpha^{\prime}l_{p}=l_{min}$, in terms of $\alpha$
 (\ref{GT3}) may be written in a natural way as follows:
\begin{equation}\label{GT5}
T_{GUP}=(\frac{d-3}{4\pi})\frac{\hbar \alpha^{-1}_{r_{+}}
}{\alpha^{\prime}l_{p}}[1-(1-\alpha_{r_{+}})^{1/2}],
\end{equation}
where $\alpha_{r_{+}}$- parameter $\alpha$ associated with the IR-cutoff
$r_{+}$. In such a manner $T_{GUP}$ is only dependent on the constants
including the fundamental ones and on the deformation parameter $\alpha$.
\\ The dependence of the black hole entropy on $\alpha$ may be derived
in a similar way. For a semiclassical approximation of the Bekenstein-Hawking formula
\cite{Bek1},\cite{Hawk1}
\begin{equation}\label{GT6}
S=\frac{1}{4}\frac{A}{l^{2}_{p}},
\end{equation}
where $A$ -- surface area of the event horizon, provided the horizon
event has radius $r_+$, then $A\sim r^{2}_+$ and (\ref{GT6}) is clearly of the form
\begin{equation}\label{GT6.1}
S=\sigma \alpha^{-1}_{r_{+}},
\end{equation}
where $\sigma$ is some dimensionless denumerable factor. The
general formula for quantum corrections \cite{mv} given as
\begin{equation}\label{GT6.2}
S =\frac{A}{4l_{p}^{2}}-{\pi\alpha^{\prime 2}\over 4}\ln
\left(\frac{A}{4l_{p}^{2}}\right) +\sum_{n=1}^{\infty}c_{n}
\left({A\over 4 l_p^2} \right)^{-n}+ \rm{const}\;,
\end{equation}
where the expansion coefficients $c_n\propto \alpha^{\prime
2(n+1)}$ can always be computed to any desired order of accuracy
\cite{mv}, may be also written as a power series in
$\alpha^{-1}_{r_{+}}$   (or  Laurent series in $\alpha_{r_{+}}$)
\begin{equation}\label{GT6.3}
S =\sigma \alpha^{-1}_{r_{+}}-{\pi\alpha^{\prime 2}\over 4}\ln
\left(\sigma \alpha^{-1}_{r_{+}}\right) +\sum_{n=1}^{\infty}c_{n}
\left(\sigma \alpha^{-1}_{r_{+}}\right)^{-n}+ \rm{const}
\end{equation}
Note that here no consideration is given to the restrictions on the IR-cutoff
\begin{equation}\label{GT7}
\textit{L}\leq \textit{L}_{max}
\end{equation}
and to those corresponding the extended uncertainty principle
(EUP) that leads to a minimal momentum \cite{Park}. This problem
will be considered separately in further publications of the
author.
\\ A black hole is a specific example of the space with horizon.
It is clear that for other horizon spaces \cite{Padm11} a similar
relationship between their thermodynamics and  the deformation
parameter $\alpha$ should be exhibited.
\\Quite recently,
in a series of papers, and specifically in
\cite{Padm3}--\cite{Padm9}, it has been shown that Einstein
equations may be derived from the surface term of the GR
Lagrangian, in fact containing the same information as the bulk
term.
\\And as Einstein-Hilbert's Lagrangian has the structure
$L_{EH}\propto R\sim (\partial g)^2+ {\partial^2g}$, in the customary
approach the surface term arising from  $L_{surf}\propto
\partial^2g$ has to be canceled to get Einstein
equations from  $L_{bulk}\propto (\partial g)^2$ \cite{Padm10}.
But due to the relationship between $L_{bulk}$ and $L_{surf}$
\cite{Padm5}--\cite{Padm7},\cite{Padm10}, we have
 \begin{equation}
\sqrt{-g}L_{suf}=-\partial_a\left(g_{ij} \frac{\partial
\sqrt{-g}L_{bulk}}{\partial(\partial_ag_{ij})}\right).
\end{equation}
In such a manner one can suggest a holographic character of
gravity in that the bulk and surface terms of the gravitational
action contain identical information. However, there is a
significant difference between the first case, when variation of
the metric $g_{ab}$ in $L_{\rm bulk}$ leads to Einstein equations,
and the second case, associated with derivation of the GR field
equations from the action principle using only the surface term
and virtual displacements of horizons \cite{Padm2}, whereas the
metric is not treated as a dynamic variable \cite{Padm10}.
\\In the case under study, it is assumed from the beginning
that we consider the spaces with horizon. It should be noted that
in the Fischler-Susskind cosmic holographic conjecture it is
implied that the Universe represents spherically symmetric
space-time, on the one hand, and has a (Hubble) horizon
(\ref{D3}), on the other hand. But proceeding from the results of
\cite{Padm3}-- \cite{Padm10}, an entropy boundary is actually
given by the surface of horizon measured in Planck's units of area
\cite{Padm6}:
\begin{equation}\label{D10}
S=\frac{1}{4}\frac{A_R}{l^{2}_{p}},
\end{equation}
where $A_R$ is the  horizon area corresponding to the Hubble
horizon $R$ (\ref{D3}).
\\To sum it up, an assumption that space-time
is spherically symmetric and has a horizon is the only natural
assumption held in the Fischler-Susskind cosmic holographic
conjecture to support its validity. Thus the arguments in support
of the Fischler-Susskind cosmic holographic conjecture are given
on the basis of the results obtained lately on Gravitational
Holography and Gravitational Thermodynamics.
\\It should be noted that Einstein's
equations may be obtained from the proportionality of the entropy
and horizon area together with the fundamental thermodynamic
relation connecting heat, entropy, and temperature \cite{Jac1}. In
fact \cite{Padm3}-- \cite{Padm10}, this approach has been extended
and complemented by the demonstration of holographicity  for the
gravitational action (see also \cite{Padm11}).And in the case of
Einstein-Hilbert gravity, it is possible to interpret Einstein's
equations as the thermodynamic identity \cite{Padm12}:
\begin{equation}\label{GT8}
TdS = dE + PdV.
\end{equation}
The above-mentioned results in the last paragraph have been
obtained at low energies, i.e. in a semiclassical approximation.
Because of this, the problem arises how these results are changed
in the case of high energies? Or more precisely, how the results
of \cite{Jac1},\cite{Padm3}-- \cite{Padm12} are generalized in the
UV-cutoff?  It is obvious that, as in this case all the
thermodynamic characteristics become dependent on the deformation
parameter $\alpha$, all the corresponding results should be
modified (deformed) to meet the following requirements:
\\(a) to be clearly dependent on the deformation parameter
$\alpha$ at high energies;
\\
\\(b) to be duplicated, with high precision, at low energies
due to the limiting transition $\alpha\rightarrow 0$.
\\
\\(c) let us clear up what is meant by the adequate $\alpha$-deformation
of Einstein's equations (General Relativity) and by the
Holographic Principle \cite{Hooft1}--\cite{Bou3}.
\\ The problem may be more specific.
\\ As, according to
\cite{Jac1},\cite{Padm11},\cite{Padm12} and some other works,
gravitation is greatly determined by thermodynamics and at high
energies the latter is a deformation of the classical
thermodynamics, it is interesting whether gravitation at high
energies (or what is the same, quantum gravity or Planck scale)is
being determined by the corresponding  deformed thermodynamics.
The formulae (\ref{GT5}) and (\ref{GT6.3}) are elements of the
high-energy $\alpha$-deformation in thermodynamics, a general
pattern of which still remains to be formed. Obviously, these
formulae should be involved in the general pattern giving better
insight into the quantum gravity, as they are applicable to black
mini-holes (Planck black holes) which may be a significant element
of such a pattern. But what about other elements of this pattern?
How can we generalize the results
\cite{Jac1},\cite{Padm11},\cite{Padm12}when the IR-cutoff tends to
the UV-cutoff (formula (\ref{GT1}))? What are modifications of the
thermodynamic identity (\ref{GT8}) in a high-energy deformed
thermodynamics and how is it applied in high-energy (quantum)
gravity? What are the aspects of using the Generalized Uncertainty
Relations in Thermodynamics
\cite{shalyt3},\cite{shalyt9},\cite{shalyt11}
(\ref{CC14a}),(\ref{CC14a})in this respect? It is clear that these
relations also form an element of high-energy thermodynamics.
\\ By authors opinion, the methods developed to solve the problem
of point (c) and elucidation of other above-mentioned problems may
form the basis for a new approach to solution of the quantum
gravity problem. And one of the keys to the {\bf quantum gravity}
problem is a better insight into the {\bf high-energy
thermodynamics}.
\section{QFT with UV-Cutoff for Different Approaches and Some Comments}
I. As shown by numerous authors (to start with \cite {Kempf}), the
Quantum Mechanics with the fundamental length (UV cutoff)
generated by GUP is in line with the following deformation of
Heisenberg algebra
 \begin{equation} \label{comm1}
 [\vec{x}, \vec{p}]=i\hbar(1+\beta^2\vec{p}^2+...)
 \end{equation}
and \begin{equation}\label{comm2} \Delta x_{\rm min}\approx
\hbar\sqrt{\beta}\sim l_{p}.
\end{equation}
In the recent works \cite {Kim2} it has been demonstrated that the
Holographic Principle is an outcome of this approach, actually
being integrated in the approach.
\\ We can easily show that the deformation parameter $\beta$
in (\ref{comm1}),(\ref{comm2}) may be expressed in terms of the
deformation parameter $\alpha$ (see Section 3 of the text)
that has been introduced in the approach associated with the density
matrix deformation. Indeed, from (\ref{comm1}),(\ref{comm2}) it
follows that $\beta\sim {\bf 1/p^{2}}$,  and for $x_{\rm min}\sim
l_{p}$, $\beta$ corresponding to $x_{\rm min}$ is nothing else but
\begin{equation}\label{comm3}
\beta\sim  1/P_{pl}^{2},
\end{equation}
where $P_{pl}$ is Planck's momentum: $P_{pl}= \hbar/l_{p}$.
\\ In this way $\beta$ is changing over the following interval:
\begin{equation}\label{comm4}
\lambda/P_{pl}^{2}\leq \beta<\infty,
\end{equation}
where $\lambda$  is  a numerical factor  and the second member in
(\ref{comm1}) is accurately reproduced in momentum representation
(up to the numerical factor) by $\alpha=l^{2}_{min}/l^{2}\sim
l^{2}_{p}/l^{2}=p^{2}/P_{pl}^{2}$
\begin{equation} \label{comm5}
[\vec{x},\vec{p}]=i\hbar(1+\beta^2\vec{p}^2+...)=i\hbar(1+a_{1}\alpha+a_{2}\alpha^{2}+...).
\end{equation}
 As indicated in the previous Section (formula (\ref{GT6.1})),
 parameter $\alpha$   has one more interesting feature:
\begin{equation} \label{comm6}
\alpha_{l}^{-1}\sim l^{2}/l_{p}^{2}\sim S_{BH}.
\end{equation}
Here $\alpha_{l}$ is the parameter $\alpha$ corresponding to $l$,
$S_{BH}$ is the black hole entropy with the characteristic linear
size $l$ (for example, in the spherically symmetric  case $l=R$ -
radius of the corresponding sphere with the surface area $A$),
and
\begin{equation} \label{comm7}
A=4\pi l^{2},S_{BH}=A/4l_{p}^{2}=\pi \alpha_{l}^{-1}.
\end{equation}
This note is devoted to the demonstration of the fact that in case
of the holographic principle validity in terms of the new
deformation parameter $\alpha$ in $QFT^{\alpha}$, considered above
and introduced  as early as 2002
\cite{shalyt12o}--\cite{shalyt12a}, all the principal values
associated with the Vacuum (Dark) Energy Problem may be defined
simply and naturally. At the same time, there is no place for such
a parameter in the well-known QFT, whereas in QFT with the
fundamental length, specifically  in $QFT^{\alpha}$, it is quite
natural \cite{shalyt1},\cite{shalyt2},\cite{shalyt4},
\cite{shalyt6},\cite{shalyt7},\cite{shalyt9}.
\\
\\II. It should be noted that
smallness of $\alpha_{R}$ (Section 3) leads to a very great
value of $r$ in (\ref{D8}),(\ref{D9}). Besides, from (\ref{D8}) it
follows that there exists some minimal entropy $S_{min} \sim S_p$,
and this is possible  only in QFT with the fundamental length.
\\
\\III.This Section is related to Section 3
in \cite{Padm1} as well as to Sections 3 and 6 in \cite{Padm2}.
The constant $l_\Lambda$ introduced in these works is such that in the
case under consideration $\Lambda\equiv l_\Lambda^{-2}$ is
equivalent to $R$, i.e. $\alpha_{R}\approx \alpha_{l_\Lambda}$
with $\alpha_{l_\Lambda}=l^{2}_{p} /l^{2}_\Lambda$. Then
expression on the right-hand side of (\ref{D6}) is the major term
of the formula for $\rho_{vac}$, and its quantum corrections are
nothing else but a series expansion in terms of
$\alpha_{l_\Lambda}$ (or $\alpha_{R}$)
\begin{equation}\label{comm8}
\rho_{\rm vac}\sim
{\frac{1}{l_p^4}\left(\frac{l_p}{l_\Lambda}\right)^2}
+{\frac{1}{l_p^4}\left(\frac{l_p}{l_\Lambda}\right)^4} + \cdots
=\alpha_{l_\Lambda} m_p^4+...
\end{equation}
 In the first variant presented in \cite{Padm1} and
\cite{Padm2} the right-hand side of (\ref{comm8}) (formulas
(12),(33)) in \cite{Padm1} and \cite{Padm2}, respectively)reveals
an enormous additional term $m_p^4\sim \rho_{QFT}$ for
renormalization. As indicated in the previous Section, it may be,
however, ignored because the gravity is described by a pure
surface term. And in the case under study, owing to the Holographic
Principle, we may proceed directly to (\ref{comm8}). Moreover, in
$QFT^{\alpha}$ there is no need in renormalization as from the
start we are concerned with the ultraviolet-finiteness.
\\ Moreover, a series expansion of (\ref{comm8}) in terms of $\alpha$
is a complete analog of the expansion in terms of the same
parameter, redetermining the measuring procedure in
$QMFL^{\alpha}$
\cite{shalyt2},\cite{shalyt4},\cite{shalyt6},\cite{shalyt9}:
\begin{equation}\label{comm9}
Sp[\rho(\alpha)]-Sp^{2}[\rho(\alpha)]=\alpha+a^{\prime}_{0}\alpha^{2}+...
\end{equation}
As indicated in \cite{shalyt10}, the same expansion may be used to
obtain quantum corrections to the semiclassical Bekenstein-Hawking
formula (\ref{D10}) for the black hole entropy.
\\IV.Besides, the Heisenberg's algebra deformations
are introduced due to the involvement of minimal length in quantum
mechanics. These deformations are stable in the sense of
\cite{Ahl}. But this is not true for the unified algebra of
Heisenberg and Poincare. This algebra does not carry the indicated
immunity. It is suggested that the Lie algebra for the interface
of the gravitational and quantum realms is in its stabilized form.
Now it is clear that such a stability should be raised to the
status of a physical principle. In a very interesting work of
Ahluwalia - Khalilova \cite{Ahl} it has been demonstrated that the
stabilized form of   the Poincare-Heisenberg  algebra
 \cite{Vilela}, \cite{Chrys} carries three additional parameters:
"a length scale pertaining to the Planck/unification scale, a
second length scale associated with cosmos, and a new
dimensionless   constant   with the immediate implication that
`point particle' ceases to be a viable physical notion. It must be
replaced by objects which carry a well-defined, representation
space dependent, minimal spatiotemporal extent".
\\Thus, within the scope of a Quantum Field Theory with the UV cutoff
(fundamental length), closeness of the theoretical and
experimental values for $\rho_{vac}$ is adequately explained. In
this case an important role is played by new parameters appearing
in the corresponding Heisenberg Algebra deformation. Specifically,
by the new small dimensionless parameter $\alpha$, in terms of
which one can adequately interpret both the smallness of
$\rho_{vac}$  and its modern experimental value. Besides, it is
shown that the Generalized Uncertainty Principle (GUP) may be an
instrument in studies of a dynamic character of the cosmological
constant $\Lambda$.
\section{Conclusion}
In conclusion it should be noted that in a series of the author's
works \cite{shalyt1}--\cite{shalyt10} a minimal
$\alpha$-deformation of QFT has been formed. By "minimal" it is
meant that no space-time noncommutativity was required, i.e. there
was no requirement for noncommutative operators associated with
different spatial coordinates
\begin{equation}\label{Concl1}
[X_{i},X_{j}]\neq 0, i\neq j.
\end{equation}
However, all the well-known deformations of QFT associated with
GUP (for example, \cite{Magg}--\cite{Kempf}) contain
(\ref{Concl1}) as an element of the corresponding deformed
Heisenberg algebra. Because of this, it is necessary to extend (or
modify) the above-mentioned minimal $\alpha$-deformation of QFT
--$QFT^{\alpha}$ \cite{shalyt1}--\cite{shalyt10} to some new
deformation $\widetilde{QFT}^{\alpha}$ compatible with GUP, as it
has been noted in \cite{shalyt13}.
\\ Besides, in this paper consideration has been given to QFT
with a minimal length, i.e. with the UV-cutoff. Consideration of
QFT with a minimal momentum (or IR-cutoff) (\ref{GT7})
necessitates an adequate extension of $\alpha$-deformation in QFT
with the introduction of new parameters significant in the
IR-limit. Proceeding from point (c) of Section 5, the problem may
be stated as follows:
\\ (c) provided $\alpha$-deformation of GR describes the ultraviolet
(quantum-gravity) limit of GR, it is interesting to examine the
deformation type describing adequately the infrared limit of GR.
It seems that some indications of a nature of such deformation may
be found from the works devoted to the infrared modification of
gravity \cite{Patil},\cite{Park1}.

\end{document}